\documentstyle[preprint,amsmath,aps,eqsecnum]{revtex}

\begin{document}

\title{Decoherence in Nanostructures and Quantum Systems}

\author{R. F. O'Connell*}
\address{Department of Physics and Astronomy, Louisiana State
University \\ Baton Rouge, LA  70803-4001}

\date{\today}
\maketitle

\begin{abstract} Decoherence phenomena are pervasive in the arena of
nanostructures but perhaps even more so in
the study of the fundamentals of quantum mechanics and quantum
computation.  Since there has been little overlap between the
studies in both arenas, this is an attempt to bridge the gap.  Topics
stressed include (a) wave packet spreading in a
dissipative environment, a key element in all arenas, (b) the
definition of a quantitative measure of decoherence, (c) the near zero
and zero temperature limit, and (d) the key role played by initial
conditions: system and environment entangled at all times so that one
must use the density matrix (or Wigner distribution) for the complete
system or initially decoupled system and environment so that use of a
reduced density matrix or reduced Wigner distribution is feasible.  Our
approach utilizes generalized quantum Langevin equations and Wigner
distributions. \\

{\noindent{PACS Codes:
03.65.Ud, 03.65.Yz, 05.40.-a, 72.15.Rn, 72.10.Bg }}

\noindent *Phone: (225)578-6848, Fax: (225)578-5855,
E-mail: rfoc@rouge.phys.lsu.edu
\end{abstract}

\newpage

\section{Introduction}

Quantum interference effects have become pervasise in two rather
disparate areas of physics, (a) mesoscopic systems
\cite{grinstein,ferry,datta,imry,bergmann,lee,chakravasty,belitz,sarma}
and (b) fundamental quantum physics  and quantum computation
\cite{bennett,quantuminfo,haroche,zeilinger,giulini,myatt,ford,ford01,ford012,ford02,murakami,fordpress1,oconnell,oconnell2,fordpress2}. 
Whereas, of necessity, there is a certain commonality of approach, in
general each area tends to emphasize the use of particular techniques.
Since our own emphasis has been in the latter area, we feel that it
might be worthwhile to present an overall account of our approach to an
audience whose main interests lie in the former area.

The generic problem of interest is the motion of a quantum particle in an
environment.  This problem has been addressed by a variety of methods
such as the Feynman-Vernon functional integral approach, the use of
master equations and various stochastic methods.  We have found that
the generalized quantum Langevin equation (GLE), supplemented by use of
the Wigner distribution function, provides the basis of a powerful and
physically transparent approach to all such problems.

In Sec. II, we discuss fundamentals.  Starting with the Hamiltonian for
the whole system, we summarize how the GLE was derived and we discuss
its properties.  The solution for a non-localized quantum particle in a
very general dissipative environment is then presented.  Results
are also presented for the correlation and
commutation of both the coordinate and the fluctuation force
operators appearing in the GLE.  In Sec. III, we derive $s(t)$, the
mean-square displacement of the quantum particle in a dissipative
environment [see (\ref{dn31}) below].  The latter result, which is
valid for arbitrary temperatures and arbitrary disspation, is a key
ingredient of a very general expression for the spreading of the width
of the particle's wave packet [see (\ref{dn311}) below].  It also plays
a vital  role in quantum transport studies.  In addition, it is relevant
to a determination of the phase-breaking time
$\tau_{\phi}$, of interest for many mesoscopic systems.  In Sec. IV, we
apply our results to a calculation of the decoherence of interference
terms in a quantum superposition.  As a measure of decoherence, we
define an attenuation coefficient $a(t)$, the general result for
which depends strongly on $s(t)$ [as may be seen from (\ref{dn43})
and (\ref{dn311}) below].  Sec. V presents a generalization to the case
of an external field and, in Sec. VI, we present our conclusions.  Fig.
1 presents a schematic summary of the progression from fundamental
theory to applications.

\section{Fundamentals}

In recent years, there has been widespread interest in dissipative
problems arising in a variety of areas in physics.  As it turns out,
solutions of many of these problems are encompassed by a generalization
of Langevin's equation to encompass quantum, memory, and non-Markovian
effects, as well as arbitrary temperature and the presence of an
external potential $V(x)$.  As in \cite{ford88}, we refer to this
as the generalized quantum Langevin equation (GLE)

\begin{equation} m \ddot{x} +\int^{t}_{-\infty}dt^{\prime}\mu
(t-t^{\prime})\dot{x}(t^{\prime})+V^{\prime}(x)=F(t)+f(t), \label{dn21}
\end{equation}  where $V^{\prime}(x)=dV(x)/dx$ is the negative of the
time-independent external force and $\mu (t)$ is the so-called memory
function.  $F(t)$ is the random (fluctuation or noise) operator force
with mean $\langle F(t)\rangle =0$ and
$f(t)$ is a $c$-number external force (due to an electric field, for
instance).  In addition, it should be strongly emphasized that "-- the
description is more general than the language --"
\cite{ford88} in that
$x(t)$ can be a generalized displacement operator (such as the phase
difference of the superconducting wave  function across a Josephson
junction).

A detailed discussion of (\ref{dn21}) appears in \cite{ford88}.  In
particular, it was pointed out that the GLE, while very general,
can be realized with a simple and convenient model, viz., the
independent-oscillator (IO) model.  The Hamiltonian of the IO system is

\begin{equation}
H=\frac{p^{2}}{2m}+V(x)+\sum_{j}\left(\frac{p^{2}_{j}}{2m_{j}}+\frac{1}{2}m_{j}\omega^{2}_{j}(q_{j}-x)^{2}
\right)-xf(t). \label{dn22}
\end{equation} Here $m$ is the mass of the quantum particle while
$m_{j}$ and
$\omega_{j}$ refer to the mass and frequency of heat-bath oscillator
$j$.  In addition, $x$ and $p$ are the coordinate and momentum
operators for the quantum particle and $q_{j}$ and $p_{j}$ are the
corresponding quantities for the heat-bath oscillators.  The infinity
of choices for the $m_{j}$ and $\omega_{j}$ give this model its great
generality\cite{ford88}.

Use of the Heisenberg equations of motion leads to the GLE (2.1)
describing the time development of the particle motion, where

\begin{equation}
\mu(t)=\sum_{j}m_{j}\omega^{2}_{j}\cos(\omega_{j}t)\theta(t),
\label{dn23}
\end{equation} is the memory function, with $\theta (t)$ the Heaviside
step function.  Also

\begin{equation} F(t)=\sum_{j}m_{j}\omega^{2}_{j}q^{h}_{j}(t),
\label{dn24}
\end{equation} is a fluctuating operator force with mean $\langle
F(t)\rangle =0$, where $q^{h}(t)$ denotes the general solution of the
homogeneous equation for the heat-bath oscillators (corresponding to no
interaction).  These results were used to obtain the results for the
(symmetric) autocorrelation and commutator of $F(t)$, viz.,

\begin{eqnarray}
\frac{1}{2}\langle F&&(t)F(t^{\prime})+F(t^{\prime})F(t)\rangle
\nonumber \\ &&~=
\frac{1}{\pi}\int^{\infty}_{0}d\omega
Re[\tilde{\mu}(\omega+i0^{+})]\hbar\omega\coth
(\hbar\omega/2kT)\cos[\omega(t-t^{\prime})] \label{dn25}
\end{eqnarray}

\begin{eqnarray} [F(t),F(t^{\prime})] &=&
\frac{2\hbar}{i\pi}\int^{\infty}_{0}d\omega Re\{\tilde{\mu}(\omega
+i0^{+})\}\omega
\nonumber \\ &&{}\times\sin\omega(t-t^{\prime}). \label{dn26}
\end{eqnarray} Here $\tilde{\mu}(z)$ is the Fourier transform of the
memory function:

\begin{equation}
\tilde{\mu}(z)=\int^{\infty}_{0}dt\mu (t)e^{izt}. \label{dn27}
\end{equation}

Equation (\ref{dn25}) is often referred to as the
second fluctuation-dissipation theorem and we note that it can be
written down explicitly once the GLE is obtained.  Also, its evaluation
requires only knowledge of Re$\tilde{\mu}(\omega)$.  On the other hand,
the first fluctuation-dissipation theorem is an equation involving the
autocorrelation of $x(t)$ and its explicit evaluation requires a
knowledge of the generalized susceptibility $\alpha(\omega)$ (to be
defined below) which is equivalent to knowing the solution to the GLE
and also requires knowledge of both Re$\tilde{\mu}(\omega)$ and
Im$\tilde{\mu}(\omega)$.  This solution is readily obtained when
$V(x)=0$, corresponding to the original Brownian motion problem
\cite{langevin}.  As shown in \cite{ford88}, a solution is also
possible in the case of an oscillator.  Taking
$V(x)=\frac{1}{2}Kx^{2}=\frac{1}{2}m\omega^{2}_{0}x^{2}$, these authors
obtained the solution of the  Langevin equation (\ref{dn21}) in the form

\begin{equation} x(t)=\int_{-\infty }^{t}dt^{\prime }G(t-t^{\prime
})\{F(t^{\prime })+f(t^{\prime})\},
\label{dn28}
\end{equation} where $G(t)$, the Green function, is given by

\begin{equation} G(t)=\frac{1}{2\pi }\int_{-\infty }^{\infty }d\omega
\alpha (\omega +i0^{+})e^{-i\omega t},  \label{dn29}
\end{equation} with $\alpha (z)$ the familiar response function

\begin{equation}
\alpha (z)=\frac{1}{-mz^{2}-iz\tilde{\mu}(z)+K}.  \label{dn210}
\end{equation} It is often convenient to write (\ref{dn28}) in the form

\begin{equation} x(t)=x_{s}(t)+x_{d}, \label{dn211}
\end{equation} where $x_{d}$ is the "driven" contribution \cite{ford00}
due to the external force $f(t)$ and $x_{s}$ is the contribution due to
the fluctuation force
$F(t)$.  Here we have introduced a subscript s to emphasize that
$x_{{\rm s}}(t)$ is a stationary operator-process, in the sense that
correlations, probability distributions, etc. for this dynamical
variable are invariant under time-translation ($t\rightarrow t+t_{0}$).
In particular, the correlation,

\begin{eqnarray} C_{o}(t-t^{\prime})\equiv\frac{1}{2}\langle x_{{\rm
s}}&&(t)x_{{\rm s}}(t^{\prime })+x_{{\rm s} }(t^{\prime })x_{{\rm
s}}(t)\rangle \nonumber \\ &&{}=\frac{\hbar }{\pi }%
\int_{0}^{\infty }d\omega {\rm Im}\{\alpha (\omega +i0^{+})\}\coth
\frac{
\hbar \omega }{2kT}\cos \omega (t-t^{\prime }),  \label{dn212}
\end{eqnarray} is a function only of the time-difference $t-t^{\prime
}$.  Furthermore, \cite{ford00}

\begin{eqnarray} C_{d}(t,t^{\prime}) &\equiv& \frac{1}{2}\langle
x(t)x(t^{\prime})+x(t^{\prime})x(t)\rangle \nonumber \\ &=&
C_{o}(t-t^{\prime})+\langle x(t)\rangle \langle x(t^{\prime})\rangle,
\label{dn213}
\end{eqnarray} where $\langle x(t)\rangle$ is the steady mean of the
driven motion.

Also, taking the Fourier transform of (\ref{dn28}), we obtain

\begin{equation}
\tilde{x}(\omega)=\alpha(\omega)\{\tilde{F}(\omega)+\tilde{f}(\omega)\},
\label{dn214}
\end{equation} where the superposed tilde is used to denote the Fourier
transform.  Thus, $\tilde{x}(\omega)$ is the Fourier transform of the
operator $x(t)$:

\begin{equation}
\tilde{x}(\omega)=\int^{\infty}_{-\infty}dtx(t)e^{i\omega t}.
\label{dn215}
\end{equation}

We have now all the tools we need to calculate observable quantities.
It is also useful to note that the commutator, which is
temperature independent, is given by the formula \cite{ford89}

\begin{equation}
\lbrack x(t_{1}),x(t_{1}+t)]=\frac{2i\hbar }{\pi }\int_{0}^{\infty
}d\omega \mathrm{Im}\{\alpha (\omega +i0^{+})\}\sin
\omega t. \label{dn216}
\end{equation}

\section{Mean Square Displacement and Wave Packet Spreading}

The mean square displacement of the quantum particle in a dissipative
environment,
$s(t)$ say, plays a key role in all of our subsequent discussions.  In
particular, it determines the diffusion time through a mesoscopic
system of interest.  In order to concentrate on the fundamental
features, we take $V=0$ and $f(t)=0$ but we will return to $f(t)\neq 0$
in Section 5.  Then using (\ref{dn212}), we obtain

\begin{eqnarray} s(t) &\equiv &\left\langle [x_{{\rm s}}(t)-x_{{\rm
s}}(0)]^{2}\right\rangle
\nonumber \\ &=& 2\left\{C_{0}(0)-C_{0}(t)\right\} \nonumber \\
&=&\frac{2\hbar }{\pi }\int_{0}^{\infty }d\omega {\rm Im}\{\alpha
(\omega +i0^{+})\}\coth \frac{\hbar \omega }{2kT}(1-\cos \omega t),
\label{dn31}
\end{eqnarray} where now

\begin{equation}
\alpha (z)=\frac{1}{-mz^{2}-iz\tilde{\mu}(z)}. \label{dn32}
\end{equation} It is informative to consider some limiting cases.  In
particular, in the quantum physics/quantum optics literature the Ohmic
model (no memory in the GLE), for which

\begin{equation}
\tilde{\mu}(\omega )=m\gamma = {\textnormal{constant
~~~~~~~~~~~~~(Ohmic)}},
\label{dn33}
\end{equation} is often considered.  Then, in the high temperature
limit one obtains
\cite{oconnell3}, with $t>0$,

\begin{equation} s(t)=\frac{2kT}{m\gamma^{2}}\left\{e^{-\gamma
t}-1+\gamma t\right\}, ~~~ kT>>\hbar\gamma. \label{dn34}
\end{equation} For long times

\begin{equation} s(t)\rightarrow \frac{2kT}{m\gamma} t,~~~\gamma t>>1
\label{dn35}
\end{equation} which is the familiar Einstein relation from Brownian
motion theory.  However, for short times (which are more characteristic
of decoherence decay times \cite{ford})

\begin{equation} s(t)\rightarrow\frac{kT}{m}t^{2},~~~\gamma t<<1,
\label{dn36}
\end{equation} independent of $\gamma$. Thus, in this simple case, the
familiar diffusion coefficient

\begin{equation}
D\equiv \frac{1}{2}\dot{s}(t)=\frac{kT}{m\gamma}, \label{dn37}
\end{equation}
so that the phase-breaking time $\tau_{\phi}\equiv\gamma^{-1}$ is given
by

\begin{equation}
\tau_{\phi}=\frac{m}{kT}D. \label{dn38}
\end{equation}
More generally, it is clear from the above that $\tau_{\phi}$ is a
complicated function of $s(t)$, depending on the temperature, on the
nature of the environment and on the time scale of interest.

In particular, the situation is very different for low temperatures
$(kT<<\hbar\gamma)$, in which case the main contribution is from the
zero-point $(T=0)$ oscillations of the electromagnetic field.  This
calculation has recently been carried out
\cite{fordpress2} and it was found necessary to use a model
incorporating memory since the Ohmic model lead to singular and
unphysical results for very short times.  In particular, the single
relaxation time $(\tau)$ model

\begin{equation}
\tilde{\mu}(z)=\frac{m\gamma}{1-z\tau}, \label{dn39}
\end{equation} leads to the result \cite{fordpress2}

\begin{equation}
s(t)=-\frac{\hbar\gamma}{\pi m}t^{2}\left\{\log\frac{\zeta
t}{m}+\gamma_{E}-\frac{3}{2}\right\},~~~
\tau <<t<<\gamma^{-1},
\label{dn310}
\end{equation} where $\gamma_{E}$ is Euler's constant.  It would be of
interest to obtain corresponding results for the actual dissipative
models which describe various mesoscopic systems.

Whereas $s(t)$ is of interest per se, it also plays a key role in the
determination of wave packet spread and attenuation of coherence.
Recently \cite{ford}, we obtained a general formula for the spreading of
a free particle Gaussian wave packet: if initially the complete system
is in equilibrium at temperature $T$ and the wave packet describing the
quantum particle has a width $\sigma$, then after a time
$t$ the width is given by

\begin{equation} w^{2}(t)=\sigma^{2}-\frac{[x(0),
x(t)]^{2}}{4\sigma^{2}}+s(t), \label{dn311}
\end{equation}
where $[x(0),x(t)]$ is the commutator.  For example, for a free
particle in a high temperature environment

\begin{equation} w^{2}(t)\rightarrow
\sigma^{2}+\frac{\hbar^{2}t^{2}}{4m^{2}\sigma^{2}}+\frac{kT}{m}t^{2}.
\label{dn312}
\end{equation}
For
$T=0$, (\ref{dn312}) reduces to the familiar formula of quantum
mechanics \cite{schiff,merzbacher}.  In general, the width $w(t)$ is a
vital ingredient in the determination of decoherence decay times, as we
shall see in the following section.

\section{Decoherence}

Decoherence refers to the destruction of a quantum interference pattern
and is relevant to the many experiments that depend on achieving and
maintaining entangled states. Examples of such efforts are in the areas
of quantum teleportation,\cite{zeilinger} quantum information and
computation,\cite{bennett,quantuminfo} entangled states,\cite{haroche}
Schr\"{o}dinger cats, and the quantum-classical
interface.  For an overview of many of the interesting
experiments involving decoherence, we refer to
\cite{haroche,zeilinger,giulini,myatt}.

Much of the discussion of decoherence
\cite{giulini,ford,ford01,ford012,ford02,murakami,fordpress1,oconnell,oconnell2,fordpress2}
has been in terms of a particle moving in one dimension that is placed
in an initial superposition state (Schr\"{o}dinger "cat" state)
corresponding to two widely separated Gaussian wave packets, each with
variance $\sigma^{2}$ and separated by a distance $d$.  For such a
state the probability distribution at time $t$ can be shown to be of
the form:

\begin{eqnarray}
P(x,t) &=&\frac{1}{2(1+e^{-d^{2}/8\sigma ^{2}})}\left\{
P_{0}(x-\frac{d}{2}
,t)+P_{0}(x+\frac{d}{2},t)\right.   \nonumber \\
&&\left. +2e^{-d^{2}/8w^{2}(t)}a(t)P_{0}(x,t)\cos \frac{[x(0),x(t)]xd}{
4i\sigma ^{2}w^{2}(t)}\right\} ,  \label{dn41}
\end{eqnarray}
where $P_{0}$ is the probability distribution for a single wave packet,
given by

\begin{equation}
P_{0}(x,t)=\frac{1}{\sqrt{2\pi w^{2}(t)}}\exp
\{-\frac{x^{2}}{2w^{2}(t)}\}.
\label{dn42}
\end{equation}
Here and in (\ref{dn41}) $w^{2}(t)$ is the variance of a single wave
packet, which in general is given by (\ref{dn311}).  Also $\sigma^{2}$
is the initial variance and $s(t)$ is given by (\ref{dn31}).  In
(\ref{dn41}) the first two terms within the braces correspond to the
two wave packets, centered at $\pm d/2$, expanding independently, while
the third term is the interference term.  Decoherence refers to the
destruction of interference, a measure of which is given by the
attenuation coefficient $a(t)$ which can be defined as the ratio of the
factor multiplying the cosine in the interference term to twice the
geometric mean of the first two terms \cite{ford,ford01}

In general, the interference term disappears in time due to either
temperature or dissipative effects.  There are various measures of
decoherence, based on decay of diagonal and off-diagonal density matrix
elements or probability distributions in phase space, momentum space or
coordinate space \cite{murakami} but we consider the latter to be the
most desirable because it is closest to experiment.  Thus, we measure
the disappearance of the interference term, that is, the loss of
coherence (decoherence), by calculating $a(t)$,
leading to the very general result \cite{ford}.

\begin{equation}
a(t)=\exp\left\{-\frac{s(t)}{8\sigma^{2}w^{2}(t)}\right\}. \label{dn43}
\end{equation}
As a concrete example, if we consider a low temperature environment
then, using (\ref{dn310}), (\ref{dn311}) and (\ref{dn43}), and
we obtain \cite{fordpress2}

\begin{equation}
a(t)=\exp\left\{\left(\frac{t}{\tau_{0}}\right)^{2}\left[\log\frac{\zeta
t}{m}+\gamma_{E}-\frac{3}{2}\right]\right\},~~~\tau <<t<<(\zeta /m)^{-1}
\label{dn44}
\end{equation}
with

\begin{equation}
\tau_{0}\equiv\frac{m\sigma^{2}}{d}\sqrt{\frac{8\pi}{\hbar\zeta}},
\label{dn45}
\end{equation}
which leads to very short decoherence times in the case of the
macroscopic separations $d$ generally considered in the quantum
physics/quantum optics arena.  However, for nanostructures, the latter
assumption is no longer valid (simply because values of $d$ are
microscopic) and, concomitantly, the short-time approximation $(\gamma
t<<1)$ will no longer be valid in many cases simply because $\tau_{d}$
values may be comparable to $\tau_{\phi}$.

\section{Presence of an external field {\lowercase{$(f(t)\neq 0)$}}}

As we saw in section 2, in particular (\ref{dn211}) and (\ref{dn213}),
the effect of an external classical force $f(t)$ is to replace
$x_{s}\rightarrow x_{s}+x_{d}$ and $C_{0}\rightarrow C_{d}$.  This
leads, with the help of (\ref{dn31}) to the conclusion that, in the
expression for $a(t)$ given by (\ref{dn44}), we simply replace
$s(t)\rightarrow s(t)+s_{d}$, where

\begin{eqnarray}
s_{d}(t) &=& \langle x^{2}_{d}(t)\rangle \nonumber \\
&=& \int^{t}_{0}dt^{\prime}\int^{t}_{0}dt^{\prime\prime}G(t-t^{\prime})
G(t-t^{\prime\prime})g(t^{\prime}-t^{\prime\prime}), \label{dn51}
\end{eqnarray}
with

\begin{equation}
g(t^{\prime}-t^{\prime\prime})=\langle
f(t^{\prime})f(t^{\prime\prime})\rangle . \label{dn52}
\end{equation}
In order to proceed further, one must, of course, specify the time
dependence of the external force.  For details, we refer to
\cite{oconnell2}.

\section{Conclusion}

The material present above is an overview of our work in the arena of
quantum physics/quantum optics, where the models we have discussed
display great generality: arbitrary temperature and arbitrary heat bath
allowing for non-Markovian (memory) effects and with no restriction to
initial decoupling of the bath and system.  They are relevant not only
to fundamental questions but also have implications for quantum
computation studies (since decoherence is always a stumbling block to
the maintenance of superposition states).  Our hope is that they will
stimulate new ideas in the study of nanostructures and mesoscopic
systems generally since analytic results have been obtained in most
cases which facilitate attainment of physical insight into the
important facets of the various problems.  The next steps should
include, in particular, an analysis of more complicated potentials and
also heat baths which display memory effects obtained from physical
models.  The latter, in turn, will have to reflect, in many cases, the
actual systems of interest since, for example, the inelastic phase
coherence time has very different forms depending on the parameters of
the system \cite{belitz,sarma}.  Furthermore as mentioned previously,
consideration of wave packets separated by nano distances will
necessitate the consideration of a greater range of decoherence decay
times with the consequence that many assumptions inherent in existing
calculations will no longer be valid.

\end{document}